\documentclass[superscriptaddress,twocolumn,showpacs,preprintnumbers,amsmath,amssymb,prb]{revtex4}

\usepackage{graphicx}
\usepackage{dcolumn}
\usepackage{bm}

\usepackage{hyperref} 
\usepackage{hypernat}

\usepackage{color} 
\definecolor{rltred}{rgb}{0.75,0,0}
\definecolor{rltgreen}{rgb}{0,0.6,0}
\definecolor{rltblue}{rgb}{0.3,0.3,1}

\hypersetup{colorlinks,%
    hypertexnames=true,%
    linkcolor=rltgreen,%
    citecolor=rltblue,%
    linktocpage=true}
    
\begin{document}
\title{Wave chaos as signature for depletion of a Bose-Einstein condensate}

\author{Iva B\v rezinov\'a}
\email{iva.brezinova@tuwien.ac.at}
\affiliation{Institute for Theoretical Physics, Vienna University of Technology,
Wiedner Hauptstra\ss e 8-10/136, 1040 Vienna, Austria, EU}  

\author{Axel U. J. Lode}
\affiliation{Theoretische Chemie, Physikalisch-Chemisches Institut, Universit\"at Heidelberg,
Im Neuenheimer Feld 229, D-69120 Heidelberg, Germany, EU}  

\author{Alexej I.~Streltsov}
\affiliation{Theoretische Chemie, Physikalisch-Chemisches Institut, Universit\"at Heidelberg,
Im Neuenheimer Feld 229, D-69120 Heidelberg, Germany, EU}  

\author{Ofir E.~Alon}
\affiliation{Department of Physics, University of Haifa at Oranim,
Tivon 36006, Israel}  

\author{Lorenz S.~Cederbaum}
\affiliation{Theoretische Chemie, Physikalisch-Chemisches Institut, Universit\"at Heidelberg,
Im Neuenheimer Feld 229, D-69120 Heidelberg, Germany, EU}  

\author{Joachim Burgd\"orfer}
\affiliation{Institute for Theoretical Physics, Vienna University of Technology,
Wiedner Hauptstra\ss e 8-10/136, 1040 Vienna, Austria, EU}  

\date{\today}

\begin{abstract}
We study the expansion of repulsively interacting Bose-Einstein condensates (BECs) in shallow one-dimensional potentials. We show for these systems that the onset of wave chaos in the Gross-Pitaevskii equation (GPE), i.e.~the onset of exponential separation in Hilbert space of two nearby condensate wave functions, can be used as indication for the onset of depletion of the BEC and the occupation of excited modes within a many-body description. Comparison between the multiconfigurational time-dependent Hartree for bosons (MCTDHB) method and the GPE reveals a close correspondence between the many-body effect of depletion and the mean-field effect of wave chaos for a wide range of single-particle external potentials. In the regime of wave chaos the GPE fails to account for the fine-scale quantum fluctuations because  many-body effects beyond the validity of the GPE are non-negligible. Surprisingly, despite the failure of the GPE to account for the depletion, coarse grained expectation values of the single-particle density such as the overall width of the atomic cloud agree very well with the many-body simulations. The time dependent depletion of the condensate could be investigated experimentally, e.g., via decay of coherence of the expanding atom cloud.
\end{abstract}
\pacs{03.75.Kk, 67.85.De, 05.60.Gg, 05.45.-a,}
\maketitle
\section{Introduction}
The workhorse for describing the non-equilibrium dynamics of Bose-Einstein condensates (BECs) of ultracold  gases is the Gross-Pitaevskii equation (GPE) (for a review see e.g.~Ref.~\onlinecite{proceed98,DalGioPitStr99}). Replacing the true many-body wave function by a single-particle orbital for the macroscopically occupied condensate (particle number $N\gg1$) results in an equation of motion that belongs to the class of nonlinear Schr\"odinger equations (NLSE). The GPE provides an appropriate starting point to investigate the underlying many-body system on the mean-field level. Effects beyond the GPE have been observed in BECs, for example in optical lattices with deep wells and small occupation numbers per site.\cite{BloDalZwe08} Other finite-number condensate effects include the demonstration of atom-number squeezing \cite{LiTucChiKas07,EstGroWel08,GroWinSchHoh09} and of Josephson junctions in a double well.\cite{GatObe07,SakStrAloCed09,SakStrAloCed10} Meanwhile, progress has been made in exploring the time-dependent many-boson Schr\"odinger equation. One approach is the multiconfigurational time-dependent Hartree for bosons (MCTDHB) method which is a numerically efficient and, in principle, exact method for the time-dependent many-body problem.\cite{StrAloCed07,AloStrCed08,mctdhb} In practice, limitations are imposed by the finite yet large number of configurations (millions) and orbitals (tens) that can be handled.\\
We investigate repulsively interacting BECs after release into shallow one-dimensional (1D) potentials. The Bose gas is dilute and, initially, practically all particles are in one single-particle state. The external potential is weak compared to the single-particle energy.
Comparison between the MCTDHB method and the GPE for the expansion of the BEC provides detailed insights to what extend the GPE is capable of describing the condensate dynamics and may be capable of mimicking excitations out of the condensate state. One case in point is our recent observation of true (physical) wave chaos in the GPE,\cite{BreColLudSchBur11} as opposed to numerical chaos \cite{HerAbl89} due to discretizations. The latter has been exploited to study e.g.~thermalization in the Bose-Hubbard system at the mean-field level.\cite{CasMasDunOls09} Two wave functions nearby in Hilbert space are exponentially separating from each other, as measured by the $L^2$ norm. Chaotic wave dynamics within the GPE is a mathematical consequence of the non-integrability resulting from the interplay between the external (one-body) potential and the nonlinearity which replaces the inter-particle interactions. Its physical implications are, however, less clear as the original many-body Schr\"odinger equation is strictly linear and, thus, regular and non-chaotic. Previously, a connection between chaotic dynamics within the GPE and growth in the number of non-condensed particles has been made for time-dependent external driving which can be seen as a source of energy.\cite{GarJakDumCirZol00} In the present study the external potentials are time independent such that the total energy is conserved. While wave chaos is likely associated with instabilities known for dynamics in periodic potentials (see e.g.~Ref.~\onlinecite{MorObe06}) it is a much more general effect since it occurs for a large class of potentials ranging from harmonic oscillators with defects to periodic and disordered potentials.
The aim of the present paper is to shed light on the physical meaning of wave chaos in the GPE for time evolution of BECs. For this purpose we compare the dynamics described by the mean-field GPE with the many-body MCTDHB method and relate the built-up of random fluctuations within the GPE to many-body observables such as the depletion of the condensate.\\
The outline of the paper is as follows. After first introducing the system under investigation in Sec.~\ref{sec:sys_inv} we briefly review the mean-field GPE and the many-body MCTDHB method and identify relevant observables (Sec.~\ref{sec:method}). The initial state whose dynamics we study upon release from the initial trapping is discussed in Sec.~\ref{sec:init_state}. We present numerical results for the dynamics in Sec.~\ref{sec:res} followed by conclusions and remarks (Sec.~\ref{sec:conc}).
\section{System under investigation}\label{sec:sys_inv}
\begin{figure}[t]
	\centering
		\includegraphics[width=9cm]{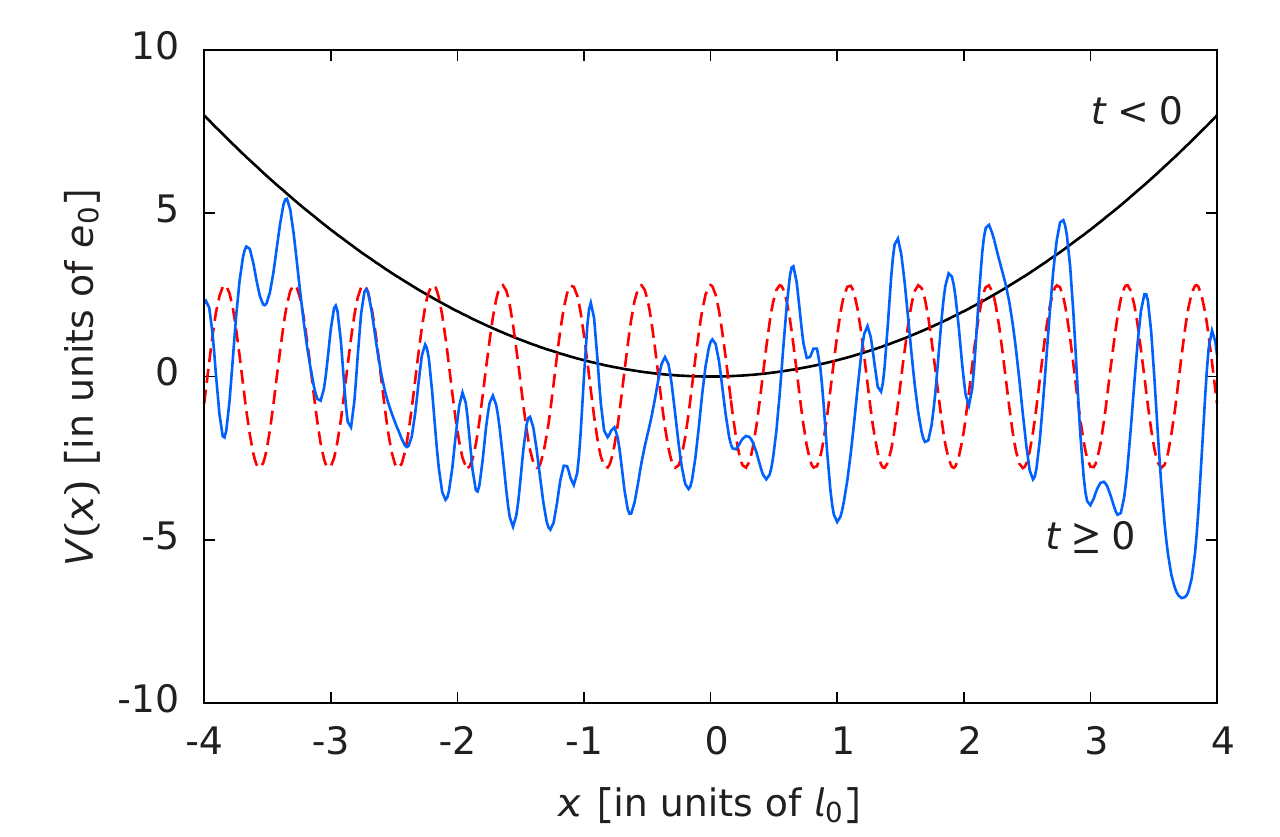}
	\caption{(Color online) The initial ($t<0$) harmonic trapping potential with longitudinal frequency $\omega_0$ (black line). At $t=0$ the longitudinal harmonic trapping potential is switched off (the radial trapping potential with frequency $\omega_r$ remains switched on). Simultaneously, either a periodic (dashed red line) or a disorder (blue solid solid) potential is switched on. The system expands for $t\geq0$ in the 1D external potential. The length scale $l_0$ corresponds to $4.6\mu$m.}
	\label{fig:potentials}
\end{figure}

We consider in the following a system of $N$ bosons interacting via a pseudo-potential which captures the scattering dynamics of the real interaction potential in the limit of small wave numbers $k\rightarrow0$. In a 1D system with tight transverse harmonic confinement with oscillator frequency $\omega_r$ the pseudo-potential is given by the contact interaction $g_{1\textrm D}\delta(x-x')$ with
\begin{equation}
g_{1\textrm D}=2\hbar\omega_ra_s,
\end{equation}
where $a_s$ is the 3D scattering length provided that $a_s\ll\sqrt{\hbar/m\omega_r}$ such that the scattering can still be regarded as a 3D process.\cite{Ols98} The dynamics of the bosonic system is then determined by the many-body Hamiltonian (in second quantization)
\begin{eqnarray}\label{eq:ham_sec}
\hat{H}&=&\int dx\; \frac{\hbar^2}{2m}\partial_x\hat{\psi}^\dagger(x,t)\partial_x\hat{\psi}(x,t) \nonumber \\
&+& \int dx\;V(x)\hat{\psi}^\dagger(x,t)\hat{\psi}(x,t) \nonumber \\
&+&\frac{g_{1\textrm D}}{2}\int dx\;\hat{\psi}^\dagger(x,t)\hat{\psi}^\dagger(x,t)\hat{\psi}(x,t)\hat{\psi}(x,t).
\end{eqnarray}
The field operators fulfill the commutation rules for bosons.
We study in the following the expansion of a Bose gas that is initially trapped also longitudinally (i.e.~in the direction of expansion) by a harmonic potential with frequency $\omega_0$ (see Fig.~\ref{fig:potentials}). These initial conditions serve to define characteristic scales for length, time, and energy. We use the units $l_0=\sqrt{\hbar/m\omega_0}$ for length, $t_0=1/\omega_0$ for time, and $e_0=\hbar\omega_0$ for energy. For a trap with $\omega_r=2\pi\times 70$Hz and $\omega_0=2\pi\times5.4$Hz used in a recent experiment on Anderson localization \cite{BilJosZuoCleSanBouyAsp08} our units take on the numerical values $l_0=4.6\mu$m and $t_0=29.47$ms.\\
We consider in the following $N=1.2\times 10^4$ $^{87}$Rb atoms. Upon release from the trap, the particles move in an external potential $V(x)$ which we specify to be a periodic potential of the form (see Fig.~\ref{fig:potentials})
\begin{equation}\label{eq:perlat}
V(x)=V_{\textrm{A}}\cos{\left( \frac{2\pi}{l}x\right)}
\end{equation}
with $l=0.54811l_0$ (corresponding to $l\approx5.8\xi$ with $\xi=\hbar/\sqrt{4m\mu}$ being the healing length and $\mu$ the chemical potential after release from the trap) and varying potential amplitude $V_{\textrm{A}}$. The periodic potential is realized in experiments by crossed laser beams in linear polarization along the same axis. For a realistic laser wave length tuned out of resonance with the $^{87}$Rb $5S\rightarrow5P$ transition, $\lambda_L=810$nm, the above potential period of $l$ corresponds to two linearly polarized crossed beams enclosing an angle of $\theta\approx0.1\pi$.\\
Alternatively, we also consider Gaussian correlated disorder potentials $V_d(x)$ of comparable strength (see Fig.~\ref{fig:potentials}). The potential is generated\cite{BreColLudSchBur11} by placing every $0.1l_0$ a Gaussian of width $\sigma$ and random weight $A_i$. The random weights are distributed uniformly in the interval $(0,1)$ (exclusive of the endpoint  values). We have used the function \texttt{ran} from Ref.~\onlinecite{NumRec} to generate the random sequences. The potential is then averaged and normalized to obtain $\langle V_d(x)\rangle=0$ and a variance of $V_{\textrm{A}}=\sqrt{\langle V_d^2(x) \rangle}$. The correlation length of the potential is $\sigma$. Unlike for the speckle potential, odd momenta $\langle V_d(x)^{2n+1}\rangle$ vanish. Moreover, the Fourier spectrum of the Gaussian correlated disorder does not have a high-momentum cutoff in contrast to the speckle potential.\cite{SanCleLugBouShlAsp07} As discussed below, our results do not display any significant qualitative difference between these two types of potentials.\\
The interplay between the inter-particle interaction and the external potential plays a key role for chaotic dynamics resulting from non-integrability.
We investigate in the following the dynamics of the expanding Bose gas in the mean-field approximation within the GPE and compare to the corresponding many-body dynamics within the MCTDHB method.
%
\section{Methods}\label{sec:method}
\subsection{Gross-Pitaevskii equation}
In the mean-field approximation the existence of a macroscopic occupation of one state is assumed such that the expectation value $\langle\hat{\psi}(x,t)\rangle=\psi(x,t)$ takes on finite values and can be treated as the classical field describing the dynamics of the BEC. Further, requiring that the expectation value of the product of four field operators factorizes
\begin{equation}\label{eq:psi4}
\langle\hat{\psi}^\dagger(x,t)\hat{\psi}^\dagger(x,t)\hat{\psi}(x,t)\hat{\psi}(x,t)\rangle=|\psi(x,t)|^4,
\end{equation}
one arrives together with 
\begin{equation}
i\hbar\frac{\partial \psi(x,t)}{\partial t}=\frac{\delta \langle \hat{H}\rangle}{\delta \psi^*(x,t)}
\end{equation}
and $\hat H$ from Eq.~\ref{eq:ham_sec} at the GPE
\begin{eqnarray}
i\hbar\frac{\partial \psi(x,t)}{\partial t}&=& -\frac{\hbar^2}{2m}\frac{\partial^2}{\partial x^2}\psi(x,t)+V(x)\psi(x,t) \nonumber \\
&+&g_{1\textrm D}|\psi(x,t)|^2\psi(x,t).
\end{eqnarray}
Normalization of the particle density to $\int dx|\psi(x,t)|^2=1$ leads to the explicit dependence of the nonlinearity on the particle number $N$:
\begin{eqnarray}\label{eq:gpe}
i\hbar\frac{\partial \psi(x,t)}{\partial t}&=& -\frac{\hbar^2}{2m}\frac{\partial^2}{\partial x^2}\psi(x,t)+V(x)\psi(x,t) \nonumber \\
&+&g_{1\textrm D}N|\psi(x,t)|^2\psi(x,t).
\end{eqnarray}
Consequently, the GPE predicts the same dynamics for different $N$ as long as the product $g_{1\textrm D}N$ is kept constant. In the limit $N\rightarrow\infty$ with $g_{1\textrm D}N=\mathrm{const.}$ the (time-independent) GPE is expected to give exact results for the many-body system (at least for the ground state of repulsive bosons in three dimensions\cite{LieSeiYng00}).\\
The parameters for the cigar-shaped trap with frequency $\omega_r$ and the particle number $N$ (see Sec.~\ref{sec:sys_inv}) together with the scattering length for $^{87}$Rb atoms\cite{BurBohEsrGre98} of $a_s\approx110a_0$ (with $a_0$ the Bohr radius) give rise to the nonlinearity
\begin{equation}\label{eq:g0}
g_0=g_{\rm {1D}}N=2\hbar\omega_ra_sN\approx390e_0l_0.
\end{equation}
Note that the rather high numerical value of $g_0$ is due to the explicit inclusion of the number of particles $N$ and does not contradict the assumption of weak interactions. Nevertheless, the interaction strength is sufficiently strong such that in the presence of an external potential depletion and fragmentation of the condensate may occur.\\
To propagate the GPE, we use a finite element discrete variable representation (DVR) to treat the spatial discretization (see e.g.~Ref.~\onlinecite{SchCol05,SchColHu06}). The propagation in time is performed by a second-order difference propagator (for details see Ref.~\onlinecite{BreColLudSchBur11} and references therein).
\subsection{Multi-configuration time-dependent Hartree for bosons (MCTDHB) method}
The MCTHDB method\cite{StrAloCed07,AloStrCed08} allows one to describe many-body effects beyond the mean-field description for the condensate. Briefly, the many-body wave function is taken as a linear combination of time-dependent permanents
\begin{eqnarray}
|\Psi(t)\rangle=\sum_{\left\{\vec{n}\right\}}C_{\vec{n}}(t)|\vec{n};t\rangle,
\label{eq:mbw}
\end{eqnarray}
where $|\vec{n};t\rangle$ corresponds to states with occupation numbers $\vec{n}=(n_1,...,n_M)$ and $M$ is the number of single-particle orbitals. The sum runs over all sets of occupation numbers $\left\{\vec{n}\right\}$ which fulfill $N=\sum_{i=1}^{M}n_i$. In the limit $M\rightarrow\infty$ the ansatz Eq.~\ref{eq:mbw} gives the exact many-body wave function. MCTDHB efficiently exploits the fact that ultracold atoms may occupy only few orbitals above the condensate state. By dynamically changing the expansion amplitudes $C_{\vec{n}}(t)$ and the orbitals $\left\{\Phi_k(\vec{r},t)\right\}$, even large many-body systems can be treated accurately. MCTDHB involves the solution of coupled linear differential equations in $C_{\vec{n}}(t)$ and coupled nonlinear differential equations in $\left\{\Phi_k(\vec{r},t)\right\}$.
The MCTDHB equations of motion reduce in the case of $M=1$ to the GPE (Eq.~\ref{eq:gpe}) with nonlinearity $g_{1D}(N-1)$. (The difference between $N$ and $N-1$ can be neglected in the limit of large $N$).\\
Within the MCTDHB method kinetic operators are treated via a fast Fourier transform which is equivalent to an exponential DVR.\cite{BecJacWorMey00} The nonlinear differential equations in the orbitals are propagated via a 5$^{th}$ order Runge-Kutta algorithm. The linear differential equations for the amplitudes $C_{\vec{n}}(t)$ are propagated via a short-iterative Lanczos algorithm. The propagations are parallelized using openMP and MPI. The Runge-Kutta algorithm has been crosschecked with the integrator\cite{BroByrHin89,Hin} ZVODE which relies on the Gear-type backwards differentiation formula for stiff ordinary differential equations and gives the same results as the faster Runge-Kutta algorithm. Further numerical checks give a very good agreement between the initial and the backwards propagated density per particle. The difference is of the order of $10^{-4}$ and less (in units of $l_0^{-1}$).
\subsection{Observables}\label{sec:observ}
The simplest and most important benchmark observable for a comparison between the mean-field and the many-body dynamics is the single-particle density. Within the MCTDHB method the density is given by 
\begin{eqnarray}\label{eq:density}
\rho(x,t)&=&\langle\hat\psi^{\dagger}(x,t)\hat\psi(x,t)\rangle\nonumber \\
&=&N\int dx_2...dx_N\;\Psi^*(x,x_2,...,x_N;t)\nonumber \\ 
&\times&\Psi(x,x_2,...,x_N;t)\nonumber \\
&=&\sum_{m,n=1}^M\rho_{m,n}(t)\Phi_m^*(x,t)\Phi_n(x,t),
\end{eqnarray}
where the elements $\rho_{m,n}(t)$ are readily accessible as a combination of the amplitudes $C_{\vec{n}}(t)$ and the corresponding occupation numbers contained in $\vec{n}$ (see Ref.~\onlinecite{AloStrCed08}). Upon diagonalization of Eq.~\ref{eq:density} the density in terms of the natural orbitals $\Phi_i^{\rm NO}(x,t)$ and their occupation numbers $n_i^{\textrm{NO}}(t)$ is obtained as:
\begin{equation}\label{eq:den_NO}
\rho(x,t)=\sum_{i=1}^{M}n_i^{\rm{NO}}(t)|\Phi_i^{\rm{NO}}(x,t)|^2.
\end{equation}
In the presence of a BEC the occupation of one state is   ``macroscopic"\cite{PenOns56} (of order $N$). In the following we denote this condensate state as $\Phi_{i=1}^{\textrm{NO}}(x,t)$ and its occupation as $n_{i=1}^{\textrm{NO}}(t)$. All other states $\Phi_i^{\textrm{NO}}(x,t)$ with $i>1$ are referred to as excited states.
The Fourier spectrum
\begin{equation}
\tilde{\rho}(k,t)=\sum_{m,n=1}^M\;\rho_{m,n}(t)\tilde\Phi_m^*(k,t)\tilde\Phi_n(k,t),
\end{equation}
is obtained by Fourier transforming the orbitals $\Phi_n(x,t)$ to give $\tilde \Phi_n(k,t)$. Within the GPE, $\tilde \rho(k,t)$ is given by the absolute square of the Fourier transform of the condensate wave function, $N|\tilde \psi(k,t)|^2$. In the limit of a long-time expansion of the BEC in free space when the initial interaction energy is converted into kinetic energy, the experimentally observed momentum distribution corresponds to the Fourier spectrum $\tilde \rho(k,t)$. \\
We utilize coherence as measured by the normalized two-particle correlation function\cite{Gla07,SakStrAloCed08} 
\begin{equation}
g^{(2)}(x_1',x_2',x_1,x_2;t) \equiv \frac{\rho^{(2)}(x_1',x_2',x_1,
x_2;t)}
{\sqrt{{\rho(x_1,t)}{\rho(x_2,t)}{\rho(x'_1,t)}{\rho(x'_2,t)}}}
\end{equation}
to analyze the breakdown of the GPE on the length scales of the random fluctuations which develop in the wave function in the regime of wave chaos. In $g^{(2)}$ the reduced two-body density matrix
\begin{eqnarray}\label{eq:rho_2}
\rho^{(2)}(x_1',x_2',x_1,x_2;t)\nonumber \\
=\langle\hat\psi^\dagger(x_1',t)\hat\psi^\dagger(x_2',t)\hat\psi(x_1,t)\hat\psi(x_2,t)\rangle\nonumber \\
=N(N-1)\int\; dx_3...dx_N\Psi^*(x_1',x_2',x_3,...,x_N;t) \nonumber \\ \times\Psi(x_1,x_2,x_3,...,x_N;t)
\end{eqnarray}
enters. For a fully second-order coherent system $g^{(2)}$ fulfills $|g^{(2)}(x_1',x_2',x_1,x_2;t)|=1$. Within the GPE the reduced two-body density matrix is a product of one-body wave functions (compare with Eq.~\ref{eq:psi4}). Thus, $|g^{(2)}|=1$ for all times, i.e., full second-order coherence is a generic feature of the GPE.
In the many-body case for a finite number of particles $N$ the departure of $|g^{(2)}|$ from $|g^{(2)}|=1-1/N$ ($|g^{(2)}|=1$ in the limit $N\rightarrow\infty$) gives a measure for how well the system is described by a single-orbital product state and how correlated
($|g^{(2)}|>1$) or anticorrelated ($|g^{(2)}|<1$) the measurement of two
coordinates is. (Anti-)Correlation indicates the degree of
fragmentation in the system.\\
As a measure for wave chaos, i.e.~the build-up of random local fluctuations on the length scale comparable to that of the external potential, we have introduced\cite{BreColLudSchBur11} the Lyapunov exponent characterizing the exponential increase of the distance in Hilbert space of two initially nearby GPE wave functions $\psi_{1,2}(x,t)$. The distance is measured by the $L^2$ norm 
\begin{eqnarray}
\label{eq:d2}
d^{(2)}(t)&=&\frac{1}{2}\int dx\;|\psi_1(x,t)-\psi_2(x,t)|^2 \nonumber \\
&=&1-\mathrm{Re}\left( \int dx\;\psi_1^*(x,t)\psi_2(x,t)\right).
\end{eqnarray}
The distance function takes on values $d^{(2)}\in[0,2]$ and is $1$ for orthogonal wave functions. In terms of $d^{(2)}$, the Lyapunov exponent which is positive in presence of chaos is given by 
\begin{eqnarray}\label{eq:lambda}
\lambda=\frac{1}{2}\lim_{t\to \infty}\lim_{d^{(2)}(0)\to 0}\frac{1}{t}\ln{\left(\frac{d^{(2)}(t)}{d^{(2)}(0)}\right)}.
\end{eqnarray}
$d^{(2)}$ is invariant for unitary time propagation of linear systems: if $\psi_1(x,t)$ and $\psi_2(x,t)$ would be solutions of the linear Schr\"odinger equation, $d^{(2)}$ would be constant. Similarly, $d^{(2)}(t)$ is constant for two many-body wave functions $\Psi_1(x_1,...,x_N,t)$ and $\Psi_2(x_1,...,x_N,t)$ integrated over all spatial coordinates. By contrast, construction of a reduced one-particle wave function from an initial $N$-body state of system 1 by (see e.g.~Ref.~\onlinecite{PinNoz99})
\begin{eqnarray}\label{eq:psi1_N}
\psi_1(x,t)&=&\langle \Psi_1(N-1)|\hat{\psi}(x,t)|\Psi_1(N)\rangle \nonumber \\ 
&=&\int dx_2...dx_N\;\Psi_1^*(x_2,...,x_N,t)\nonumber \\
&&\times\Psi_1(x,x_2,...,x_N,t),
\end{eqnarray}
where $|\Psi_1(N)\rangle$ denotes a many-body state with $N$ particles leads to a many-body measure analog to the $d^{(2)}(t)$ function that is not conserved as a function of time. This can be seen by inserting Eq.~\ref{eq:psi1_N} into Eq.~\ref{eq:d2} and taking the time derivative of $d^{(2)}$. In the time derivative of $d^{(2)}$ contributions originating from the kinetic energy and the external potential $V(x)$ cancel, while contributions from the interaction term lead to $d [d^{(2)}(t)]/dt\neq0$. The tracing out of unobserved degrees of freedom leads to the violation of the distance conserving evolution. In the case of the GPE the nonlinearity present can cause exponential divergence of $d^{(2)}$.\\
It is now our aim to relate the behavior of the $d^{(2)}$ function within the GPE to properties of the time evolution of the underlying many-body system. The working hypothesis is that the random fluctuations developing within the GPE are the signature for its failure to properly account for the depletion of the condensate, i.e.~excitation of the BEC during expansion in an external potential. In turn, within the MCTDHB method the population of all natural orbitals beyond that describing the condensate should grow. While at $t=0$ the MCTDHB method and the GPE closely agree which each other with only one natural orbital occupied, $\bar n_1^{\textrm{NO}}(0)=n_1^{\textrm{NO}}(0)/N\approx1$ (see next Sec.~\ref{sec:init_state}), with increasing time all other occupation numbers $\bar n_i^{\textrm{NO}}$ $(i>1)$ should increase. In the following we study the dynamics of $N=10^3$ to $10^5$ particles for which the ground state densities closely agree with each other (see Fig.~\ref{fig:gs_density}). For $N=10^4$ and $N=10^5$ only $M=2$ orbitals allow a numerically feasible number of configurations ($N_c=10^4 +1$ to $N_c=10^5+1$, respectively). Already adding one more orbital ($M=3$) leads to a configuration size of $N_c=50,015,001$ for $N=10^4$ which may be at the border of feasibility and requires a massive parallelization over a large number of processors. The system with $N=10^5$ and $M=3$ resulting in $N_c\approx5\times 10^9$ is out of reach for the current implementation of the MCTDHB method. For $N=10^3$ a number of orbitals up to $M=3$ is numerically feasible and allows to quantify the effect of adding one more orbital to the case $M=2$. Due to the numerical limitations we focus on the early stages of the depletion process when the depletion is still relatively weak $\bar n_1^{\textrm{NO}}(t)\geq0.95$.\\
As a measure for the depletion we introduce the state entropy for a general many-body state
\begin{equation}\label{eq:entropy}
S_{N}(t)=-\sum_i\bar n_i^{\textrm{NO}}(t)\ln \bar n_i^{\textrm{NO}}(t),
\end{equation}
where $\bar n_i^{\rm NO}(t)=n_i^{\rm NO}(t)/N$. For the initial conditions used in the present study we have $S_N(t)\geq S_N(0)\gtrsim0$. Note that for finite $N$, $S_N(0)$ is not exactly zero for the interacting ground state since the condensation is not complete (see Sec.~\ref{sec:init_state}). Within the GPE, where $n_1^{\rm NO}(t)=1$ and $n_{i>1}^{\rm NO}(t)=0$, $S_N(t)$ remains strictly zero. Deviations of $S_N(t)$ from zero within the many-body theory thus mark deviations from the GPE. In the following we will focus on the time evolution of Eq.~\ref{eq:entropy} and investigate the time scale of depletion, $t_d$, defined by the occurrence of an abrupt change of $S_N$ from $S_N\approx0$ to $S_N>0$. We associate this quantity with the onset of exponential growth, $t_e$, of $d^{(2)}(t)$ within the GPE. $t_e$ is determined from the crossing point between the free-space expansion behavior of $d^{(2)}(t)$ (in Fig.~\ref{fig:te} dashed curve) and the exponential fit to the increase in presence of an external potential (in Fig.~\ref{fig:te} dotted curve). $t_d$ is implicitly dependent on $N$ through the degree of coherence of the condensate. The larger $N$, the smaller the depletion ($\bar n_i^{\textrm{NO}}$, $i\ge2$) at the same time. Consequently, the depletion time is size dependent $t_d(N)$ (see below).
\section{The initial state}\label{sec:init_state}
\begin{figure}[t]
	\centering
		\includegraphics[width=9cm]{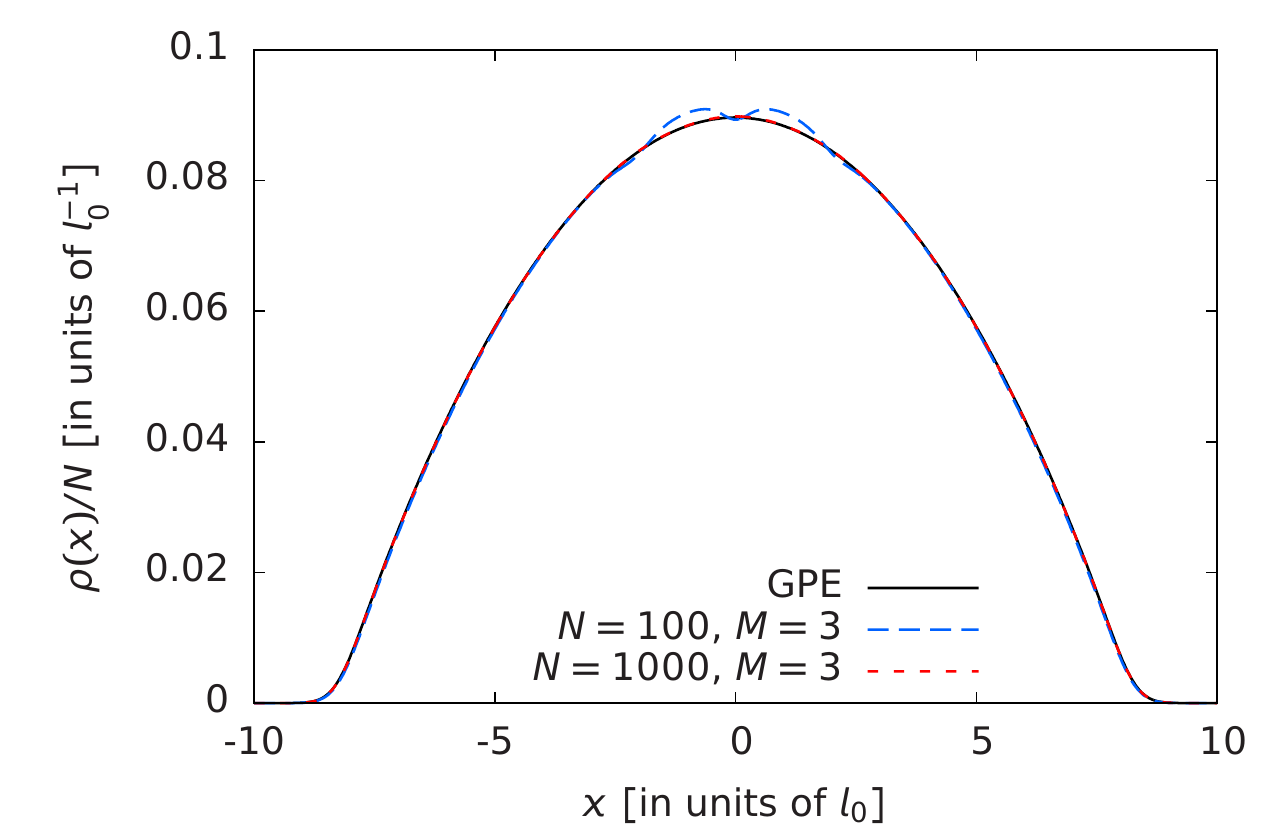}
	\caption{(Color online) The initial state of the BEC in the harmonic trap with $\omega_0=2\pi\times5.4$Hz and $l_0\approx4.6\mu$m. The interaction strength is given by the nonlinearity $g_0$ (Eq.~\ref{eq:g0}). Results for GPE (black line) and MCTDHB with N=1000, M=3 (red short dashed line) are indistinguishable within the graphical resolution; blue dashed line, MCTDHB with $N=100$, $M=3$. The two local maxima for $N=100$ are due to depletion of the condensate in the initial state and indicate deviations from the Thomas-Fermi limit.}
	\label{fig:gs_density}
\end{figure}

The initial state of the bosonic gas corresponds to the ground state of the harmonic trap. For this ground state the GPE with nonlinearity $g_0$ predicts a BEC in the Thomas-Fermi regime. Applying both the MCTDHB method and the GPE to the same system requires a careful choice of system parameters, in particular the particle number $N$. While the validity of the GPE calls for the limit of large $N\rightarrow\infty$, such a case is numerically prohibitive for the MCTDHB expansion Eq.~\ref{eq:mbw}. Since the GPE results are invariant for varying $N$ but fixed $g_0=g_{1\textrm D}N$ we adjust the particle number such as to remain in the Thomas-Fermi limit of the longitudinally trapped BEC (see Fig.~\ref{fig:gs_density}). In
such a way it is assured that discrepancies between the GPE and the MCTDHB method during the time evolution are not
caused by incompatible initial conditions.\\
The ground state of an interacting system of bosons trapped by a harmonic potential is governed by three length scales: the characteristic length of the harmonic trap $l_0$, the mean inter-particle distance $r_s=n^{-1}$ with $n$ the particle number per unit length, and 
$l_{\delta}=\frac{\hbar^2}{mg_{1\textrm D}}$ a measure for the zero-point fluctuations (or anti-correlation length) of the repulsive two-body delta-function interactions of strength $g_{1\textrm D}$. The regimes obtained range from a non-interacting ``Gaussian" shaped BEC, over a Thomas-Fermi BEC, to a strongly interacting fermionized Tonks-Girardeau gas.\cite{PetShlWal00} The presence or absence of a BEC is determined by the ratio
\begin{equation}\label{eq:gamma}
\gamma=\frac{r_s}{l_\delta},
\end{equation}
referred to as the Lieb-Lininger parameter.\cite{LieLin63} If $l_\delta$ is much larger than the inter-particle spacing $r_s$ the particles favor to occupy the same state and form a BEC. The condition for the presence of a BEC thus is:
\begin{equation}\label{eq:cond_bec}
\gamma\ll1.
\end{equation}
In order to distinguish between a Gaussian and a Thomas-Fermi BEC the harmonic oscillator length $l_0$ must be considered. The regimes are controlled by the parameter\cite{PetShlWal00}
\begin{equation}\label{eq:alpha}
\alpha=\frac{l_0}{l_\delta}=\frac{g_{1\textrm D}}{e_0l_0}.
\end{equation}
In our case $\alpha\ll1$ ($\alpha\approx390/N$ with the numerical value from Eq.~\ref{eq:g0}), i.e.~$l_0\ll l_\delta$. If in addition $l_0\gg r_s$ the system is in the Thomas-Fermi regime. The condition $l_0\gg r_s$ implies $N\gg\alpha^{-1}$ for the Thomas-Fermi limit to hold.\cite{PetShlWal00} For all systems with $N\ge1000$ in Tab.~\ref{tab:sys} the criteria $N\gg\alpha^{-1}$ and $\gamma\ll1$ are well fulfilled and, indeed, the many-body ground state density takes on the Thomas-Fermi shape (see Fig.~\ref{fig:gs_density}). The density is practically indistinguishable from the GPE prediction. For comparison, we also show in Fig.~\ref{fig:gs_density} a system with $N=100$ for which the criterion of a Thomas-Fermi BEC is only marginally fulfilled because $\gamma\approx1$ and deviations become apparent.\\
\begin{table}[b]
\centering
\begin{tabular*}{0.4\textwidth}{@{\extracolsep{\fill}} l  l  l  l  l }
\hline
$N$     &  $M$     &  $\alpha=g_{1\textrm D}[e_0l_0]$    &    $n_1^{\textrm{NO}}/N$        &   $n_M^{\textrm{NO}}/N$              \\ \hline\hline
$10^3$  &  $3$     &   $0.39$     &    $0.995$        &   $0.205\times10^{-2}$  \\ 
$10^3$  &  $2$     &   $0.39$     &    $0.997$        &   $0.266\times10^{-2}$  \\ 
$10^4$  &  $2$     &   $0.039$    &    $0.9997$       &   $0.33\times10^{-3}$   \\ 
$10^5$  &  $2$     &   $0.0039$   &    $0.99997$      &   $0.34\times10^{-4}$   \\ \hline
\end{tabular*}
\caption{Parameters of the many-body systems trapped in the harmonic oscillator at $t=0$ (Fig.~\ref{fig:gs_density}): Particle number $N$, number of orbitals $M$. The interaction strengths $g_{1\textrm D}$ correspond to constant $g_0=g_{1\textrm D}N$ (Eq.~\ref{eq:g0}). For the definition of $\alpha$ see Eq.~\ref{eq:alpha}. Highest occupation number $n_1^{\textrm{NO}}$, smallest occupation number $n_M^{\textrm{NO}}$. The parameter $\gamma$ (Eq.~\ref{eq:gamma}) fulfills $\gamma\ll1$.}\label{tab:sys}
\end{table}
The requirement of large N places a severe limit on the number of orbitals that allow for a numerically feasible configuration space. Convergence in the orbital number is controlled by the occupancy $n_M^{\textrm{NO}}$ of the least occupied state. While for the ground state calculations $n_M^{\textrm{NO}}$ is sufficiently low, we expect this number to rapidly increase during expansion since strong depletion may occur.
We, therefore, expect only the onset of depletion to be quantitatively reliable while the occupation numbers of excited orbitals can be considered to be an indication of the excitation process as the orbital expansion ceases to converge ($M>3$ time-dependent orbitals would be needed) with increasing propagation time.
\section{Numerical results}\label{sec:res}
%
\begin{figure}[t]
	\centering
		\includegraphics[width=9cm]{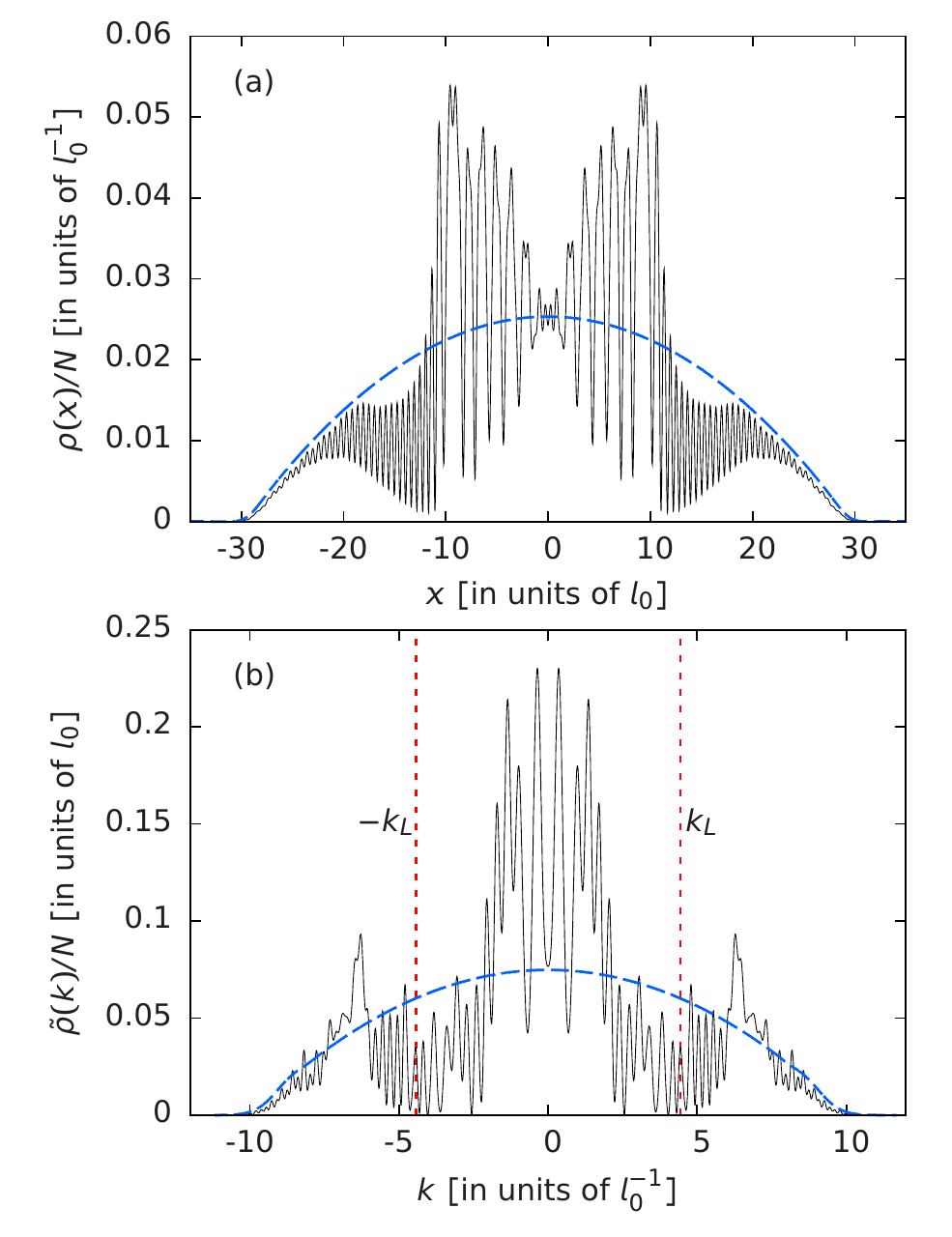}
	\caption{(Color online) (a) The density of an expanding cold atom cloud in a periodic potential within the MCDTHB for $N=10^5$ at $t=3t_0$. The inverted parabola in blue dashed line indicates the shape of an expanding cloud in the absence of the potential. The parameters of the periodic potential Eq.~\ref{eq:perlat} are $l=0.54811 l_0$ and $V_{\textrm{A}}=0.2e$, where $e$ is the energy per particle on the mean-field level. (b) The corresponding Fourier spectrum again compared to its form for free expansion (blue long dashed parabolic curve). The vertical red short dotted lines correspond to $\pm k_L=\pm4.434l_0^{-1}$, the momenta assiociated with the Landau velocity (see text).}
	\label{fig:rho_ftrho_t3}
\end{figure}
\begin{figure}[t]
	\centering
		\includegraphics[width=9cm]{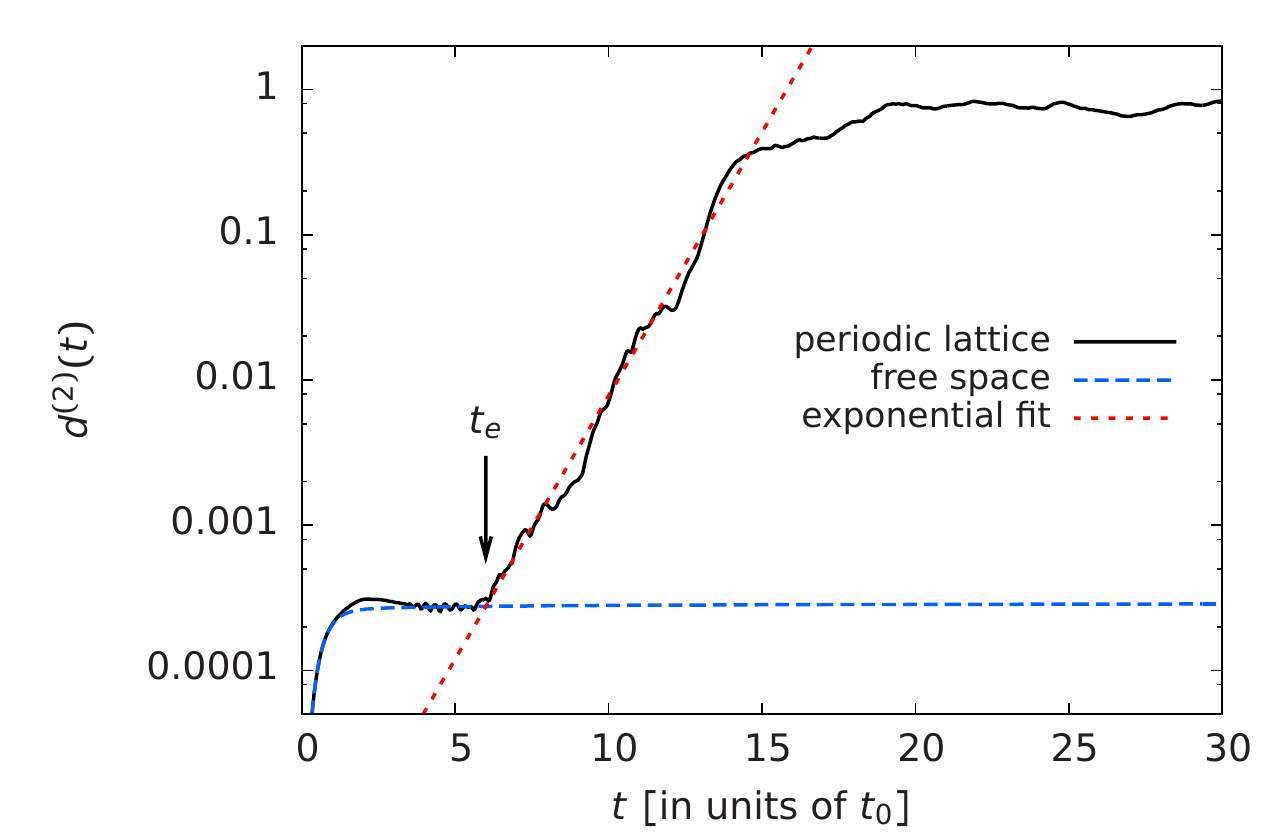}
	\caption{(Color online) Time dependence of the distance function $d^{(2)}(t)$ within the GPE for a periodic potential with period $l=0.54811l_0$ and amplitude $V_{\rm A}=0.2e$ (black solid line) as well as for free space expansion (blue dashed line) after release from the harmonic trap. The initial wave functions $\psi_1(x,0)$ and $\psi_2(x,0)$ correspond to weakly perturbed ground state wave functions given in Fig.~\ref{fig:gs_density} with $d^{(2)}(0)\approx10^{-7}$ (for details see Ref.~\onlinecite{BreColLudSchBur11}). The nonlinearity is $g_0\approx390e_0l_0$ (Eq.~\ref{eq:g0}). $d^{(2)}(t)$ for the periodic potential is fitted to an exponential function (red short dashed line). The time $t_e$ is determined from the crossing point between the exponential and the free space expansion. The saturation of $d^{(2)}(t)$ for $t\gtrsim15t_0$ near unity indicates approximate orthogonality of $\psi_1(x,t)$ and $\psi_2(x,t)$. }
	\label{fig:te}
\end{figure}
\begin{figure}[t]
	\centering
		\includegraphics[width=8.5cm]{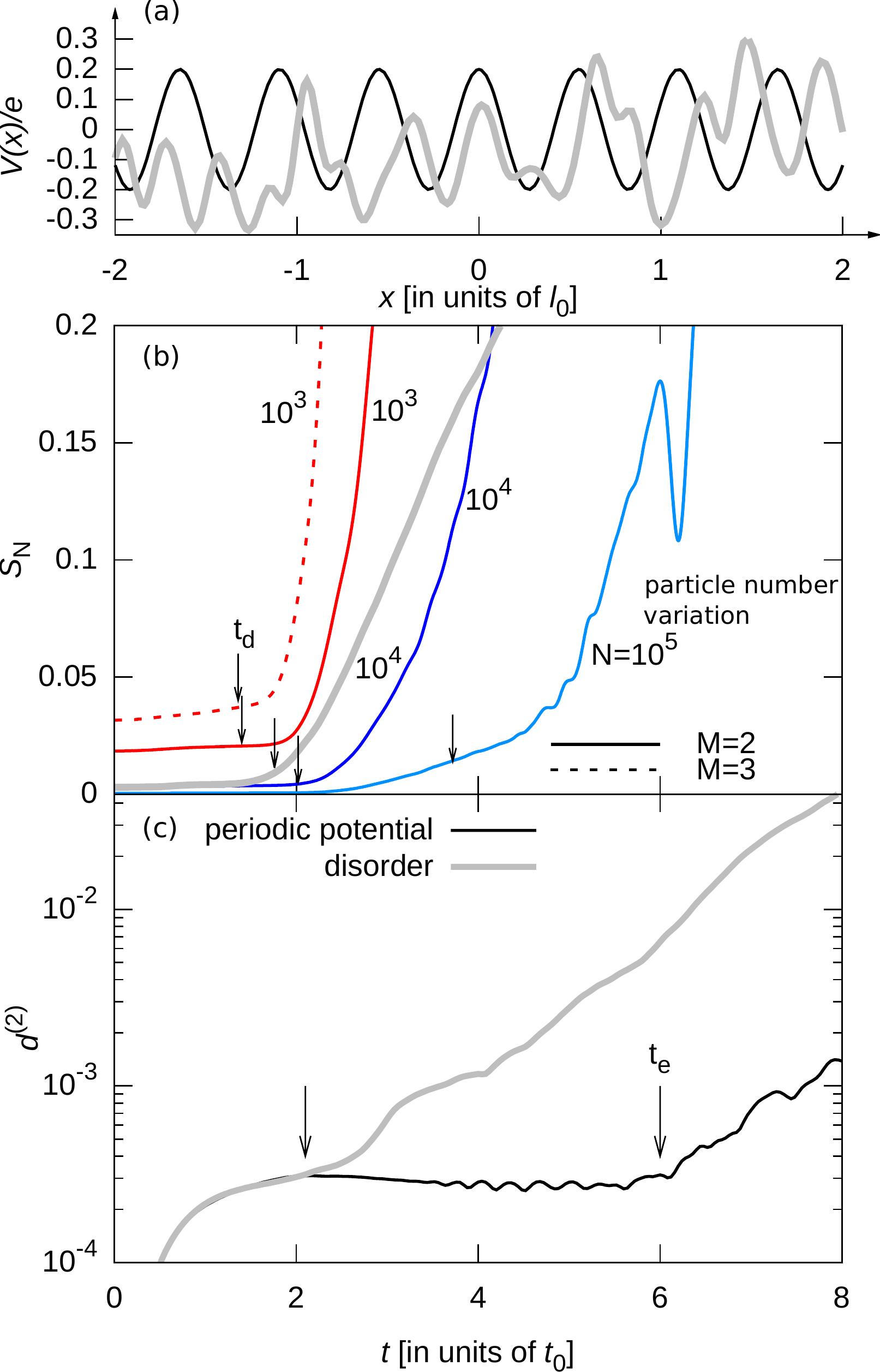}
	\caption{(Color online) (a) The periodic and a sample realization of the disorder potential with amplitude $V_{\textrm{A}}=0.2e$. The period of the periodic potential is $l=0.54811l_0\approx5.8\xi$. The correlation length of the disorder potential is $\sigma=0.7\xi$. (b) Onset of depletion within the MCTDHB for different particle and orbital numbers for propagation in the periodic and disorder potential of (a). The numbers next to the curves indicate the particle number $N$. (c) Onset of chaos within the GPE. In (a), (b), and (c) the thick light gray line corresponds to the disorder potential. Within the GPE $d^{(2)}$ has been determined by averaging over 90 realizations of the disorder potential. Within the MCTDHB 44 realizations have been used for $S_N(t)$. The onset of depletion $t_d$ and exponential divergence $t_e$ are marked by arrows.}
	\label{fig:sn_d2_Npart}
\end{figure}
\begin{figure}[tb]
	\centering
		\includegraphics[width=8.5cm]{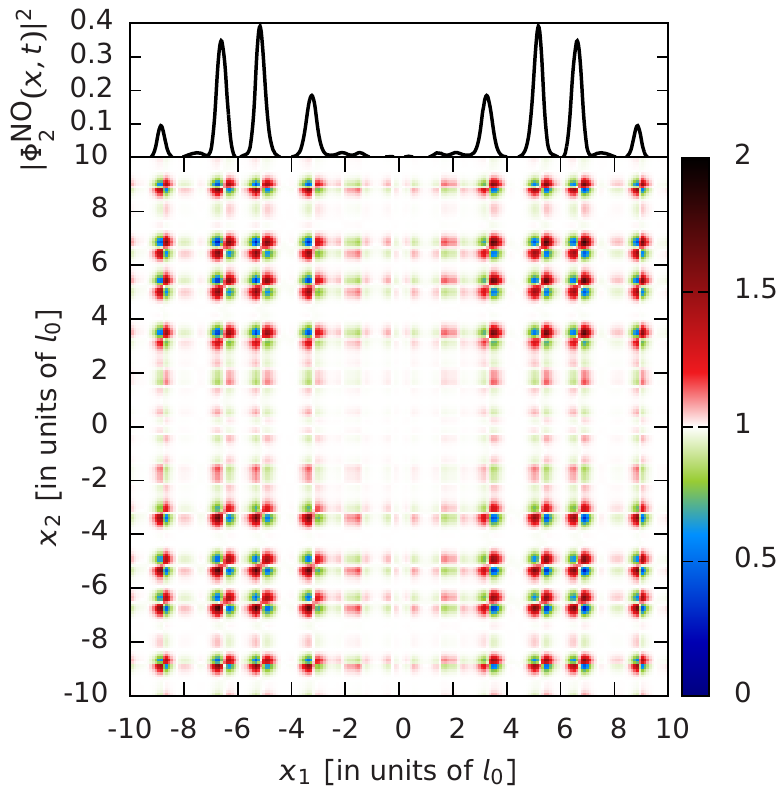}
	\caption{(Color online) Two-particle normalized correlation function $g^{(2)}
(x_1,x_2,x_1,x_2;t=4t_0)$ for $N=10^4$ bosons
as well as the density of the second natural orbital $|\phi^{\textrm{NO}}_2(x,t=4t_0)|^2$
(upper frame)
obtained from MCTDHB. The color code is chosen to highlight deviations from full coherence (white): red corresponds to correlations ($g^{(2)}>1$) and blue to anti-correlations ($g^{(2)}<1$). Note that $g^{(2)}(x_1,x_2,x_1,x_2;t)$ is a real function (see Eq.~\ref{eq:rho_2}) and is equal to unity within the GPE. For a movie of the time dependence of $g^{(2)}(x_1,x_2,x_1,x_2;t)$ as well as $n_2^{\rm NO}(t)|\phi^{\textrm{NO}}_2(x,t)|^2$ see Ref.~\onlinecite{Sup}.}
	\label{fig:g2_N104}
\end{figure}
\begin{figure}[t]
	\centering
		\includegraphics[width=8.5cm]{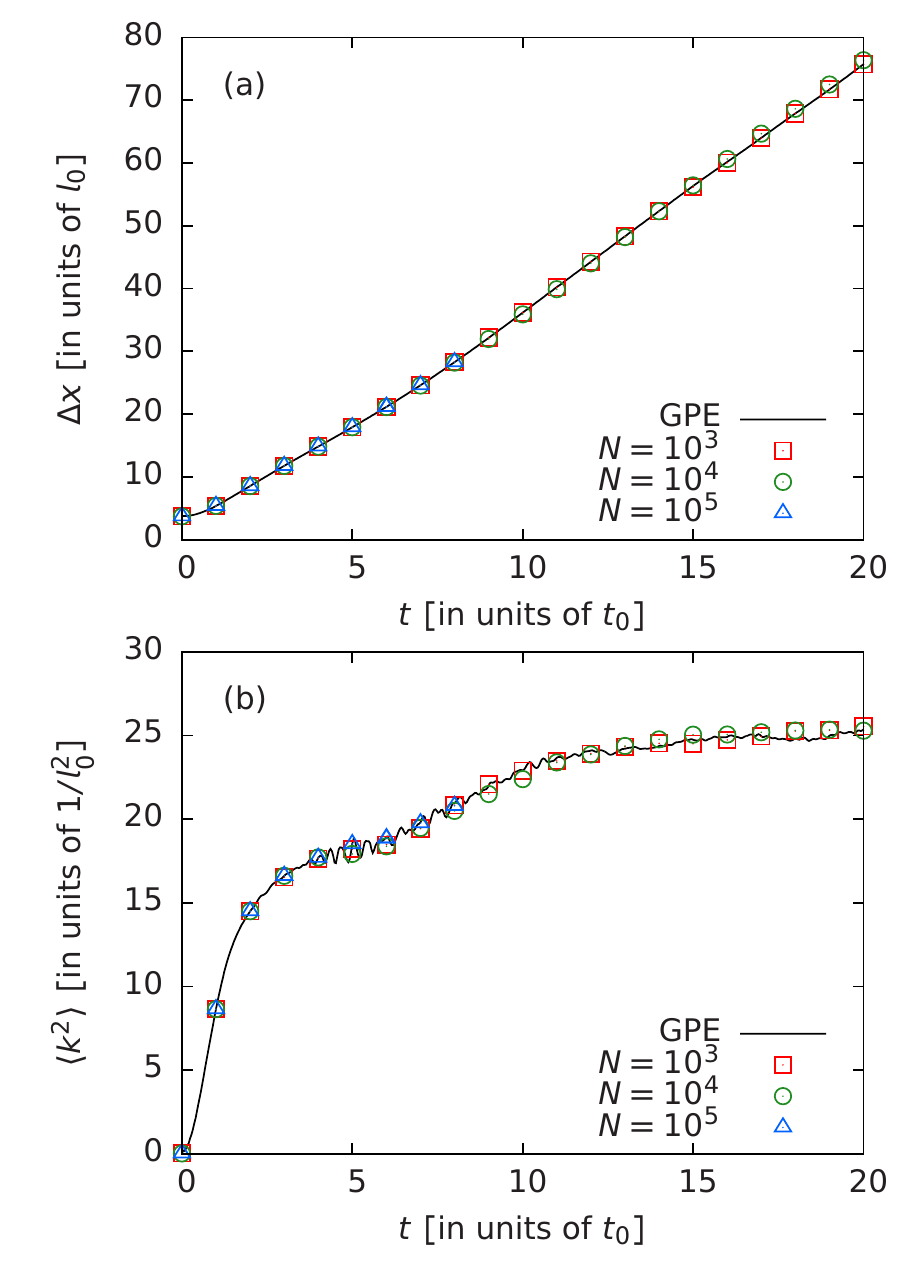}
	\caption{(Color online) (a) The width $\Delta x$ as a function of time as predicted by the GPE (solid black line) and the MCTDHB for $N=10^3$ (red squares), $N=10^4$ (green circles), and $N=10^5$ (blue triangles). (b) Average of the square of momentum $\langle k^2 \rangle$ (or mean kinetic energy) as a function of time, symbols as in (a).}
	\label{fig:xn_kn_average}
\end{figure}

We first consider the expansion of the BEC which is initially formed inside the harmonic trap (Fig.~\ref{fig:gs_density}) and then released into a periodic potential (Eq.~\ref{eq:perlat}) with $l=0.54811 l_0$ ($l\approx5.8\xi$) and $V_{\textrm{A}}=0.2e$ with $e$ the total energy per particle. After the release an explosion-like process takes place: the interaction energy is rapidly transformed into kinetic energy. In free space the cloud expands keeping its Thomas-Fermi shape with the characteristic length increasing in time.\cite{KagSurShl97} This process is modified by the presence of the periodic potential. Practically immediately the density is modulated by standing waves with the same spatial periodicity as the potential. The local maxima of the density coincide with the local minima of the potential and lead to an increase of kinetic and interaction energy at cost of potential energy.\\
As soon as the Fourier spectrum is sufficiently broad, inelastic processes set in. As momenta increase to $k\simeq k_L=\frac{mv_L}{\hbar}$ with $v_L$ the Landau velocity, the threshold for excitation of phonons, i.e.~friction of superfluid flow is reached. For a homogeneous system $v_L$ is given by $v_L=\sqrt{\frac{\mu}{m}}$ with $\mu=ng_{\textrm{1D}}$ and $n$ the particle density. By applying this relation with $\mu$ from the inhomogeneous system we determine $k_L$ from $v_L=\sqrt{\frac{\mu}{m}}$. At $t\approx3t_0$ the width $\Delta k$ of the Fourier spectrum is approximately as large as $k_L$ and we observe the development of strong density modulations [Fig.~\ref{fig:rho_ftrho_t3} (a)]. These spatial density modulations go hand in hand with reduced density in the Fourier spectrum near $\pm k_L$ since those particles lose their momentum by phonon excitations [Fig.~\ref{fig:rho_ftrho_t3} (b)]. Friction leads to the separation of a strongly fluctuating central part of the density from its fast tails [Fig.~\ref{fig:rho_ftrho_t3} (a)]. The tails expand nearly freely and are modulated by the potential. We point out that this process is fully accounted for within the GPE (i.e., the system remains condensed) since it gives practically the same density and spectrum for $t=3t_0$ as MCTDHB in Fig.~\ref{fig:rho_ftrho_t3}.\\
For longer times we have previously observed for this system signatures of wave chaos:\cite{BreColLudSchBur11} two nearby effective one-body wave functions $\psi_1(x,t)$ and $\psi_2(x,t)$ (with initially large overlap) propagated by the GPE become orthogonal to each other after an exponential increase in distance in Hilbert space (see Fig.~\ref{fig:te}). The exponential increase sets in at a characteristic time $t_e$ subsequent to a universal (i.e.~independent of the external potential) increase of $d^{(2)}(t)$ for times $t\lesssim2t_0$ (see Ref.~\onlinecite{BreColLudSchBur11} and Fig.~\ref{fig:te}). We fit the increase of $d^{(2)}(t)$ to an exponential with the Lyapunov exponent $\lambda$ as the slope (see Eq.~\ref{eq:lambda}). As soon as $d^{(2)}(t)$ reaches $d^{(2)}(t)\approx1$ the curve saturates because orthogonality, i.e.~the maximal distance in Hilbert space, is reached. Orthogonality results from the build-up of random local fluctuations in the wave functions on length scales comparable to the period of the potential.\\
We now compare the growth in $d^{(2)}(t)$ within the mean-field description with the growth of $S_N(t)$ (or depletion) within the MCTDHB method which the GPE cannot represent.
\begin{figure}[t]
	\centering
		\includegraphics[width=8.5cm]{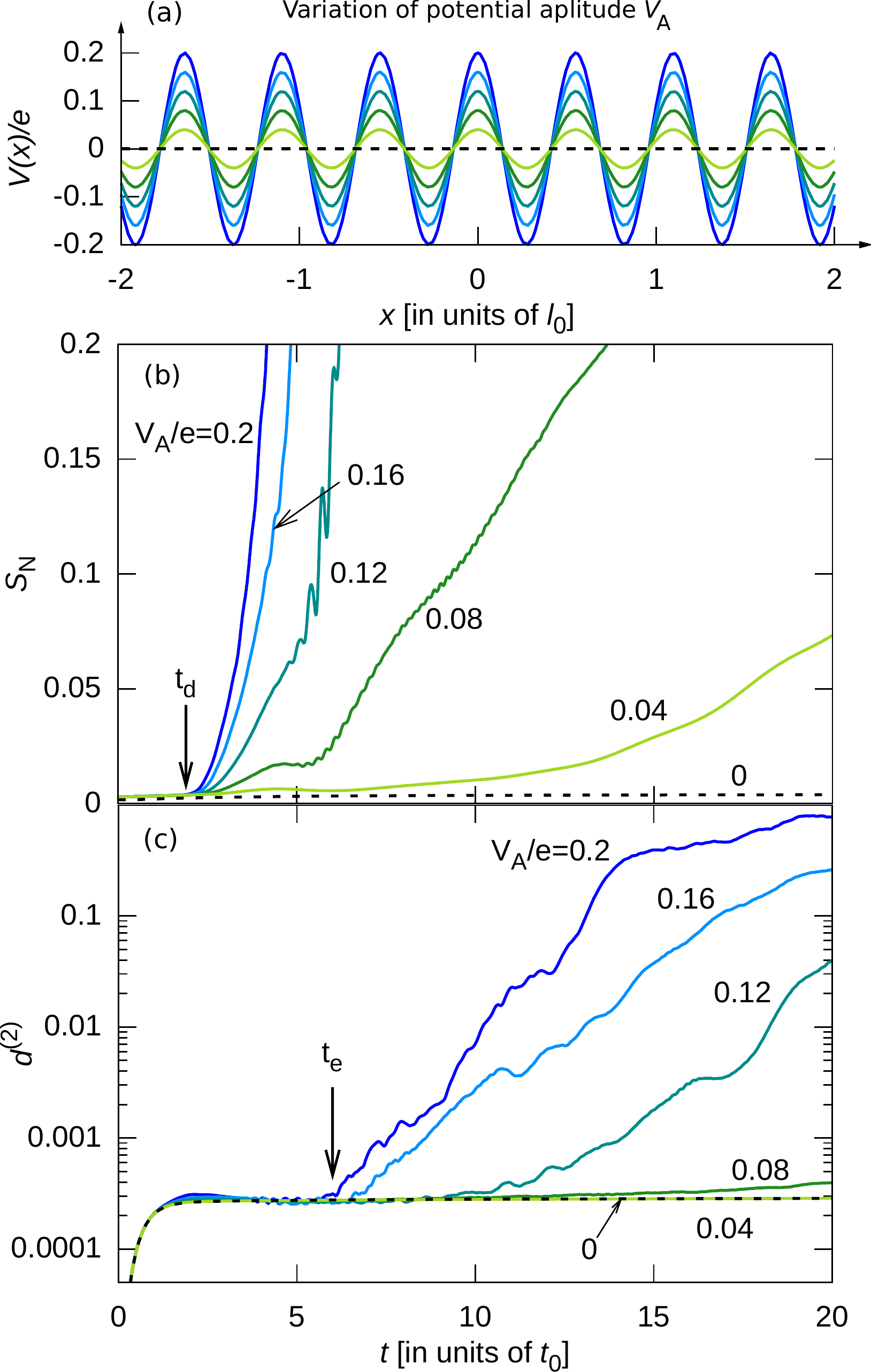}
	\caption{(Color online) (a) The potential $V(x)$ in units of the energy per particle $e$ for varying amplitude $V_{\textrm{A}}$, the period $l$ is $l=0.54811l_0$. (b) The state entropy $S_N(t)$ within MCTDHB and (c) the $d^{(2)}(t)$ function of two close wave functions within the GPE expanding in the periodic potentials of (a). In (b) and (c) $V_{\textrm{A}}$ in units of the energy per particle $e$ is indicated next to the curves. The black dashed line in (a), (b), and (c) refers to vanishing potential. For the many-body system the particle number is $N=10^4$, the orbital number is $M=2$. The onset of exponential growth $t_e$ as well as the onset of depletion $t_d$ for $V_{\textrm{A}}=0.2e$ are marked by arrows. $t_d$ corresponds to a time of $\approx60$ms. Note that $S_N$ is strictly zero within the GPE.}
	\label{fig:sn_d2_threshold}
\end{figure}
\begin{figure}[t]
	\centering
		\includegraphics[width=8.5cm]{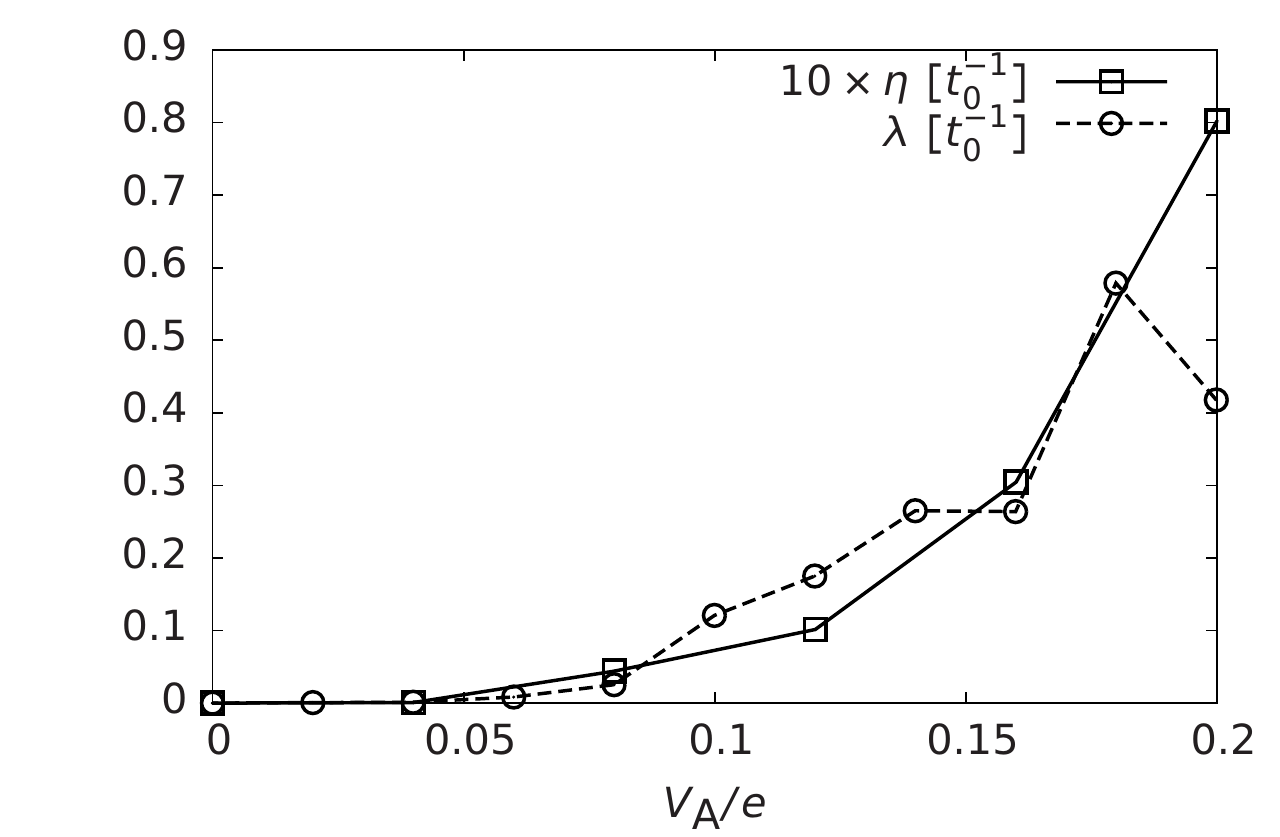}
	\caption{The depletion rate $\eta$ as a function of $V_{\textrm{A}}$ as compared to the Lyapunov exponent $\lambda$. $\eta$ and $\lambda$ are calculated from $S_N(t)$ and $d^{(2)}(t)$, respectively. The MCDTHB calculation is for $N=10^4$ particles. All other parameters as in Fig.~\ref{fig:sn_d2_threshold}.}
	\label{fig:lambda_eta}
\end{figure}
For vanishing potential $V(x)=0$ we find that the explosion-like expansion with a rapid transformation of interaction energy to kinetic energy does not lead to depletion of the condensate [Fig.~\ref{fig:sn_d2_threshold} (b) dashed line]. The GPE accounts for the expansion dynamics since $S_N(t)$ remains approximately zero as a function of time. For vanishing potential the GPE is integrable\cite{DraJoh96} such that $d^{(2)}(t)$ saturates after a short universal increase [Fig.~\ref{fig:te} and Fig.~\ref{fig:sn_d2_threshold} (c) dashed line]. For periodic potentials we find a drastic increase of $S_N(t)$ within the MCTDHB as a function of time (Fig.~\ref{fig:sn_d2_threshold}) mirroring the exponential increase in $d^{(2)}(t)$ within the GPE. To extract the rate of depletion $\eta$ and the depletion time $t_d$ we fit $S_N(t)$ to functions of the form
\begin{equation}\label{eq:fit}
S^f_N(t)=S_N(0)+c\left(\frac{t}{t_0}-\frac{t_d}{t_0}\right)^{a}\Theta\left(\frac{t}{t_0}-\frac{t_d}{t_0}\right)
\end{equation}
with fit parameters $c$, $a$, and $t_d$ (in Fig.~\ref{fig:sn_d2_threshold} $t_d$ within the MCTDHB is marked for $V_{\textrm{A}}=0.2e$). $\Theta$ is the Heaviside step function. We introduce the depletion rate $\eta$ as 
\begin{equation}
\eta=\frac{c a}{t_0}.
\end{equation}
The depletion rate $\eta$ is equal to the slope of $S^f_N(t)$ at $t=t_d+t_0$, i.e.~after the abrupt increase of $S^f_N(t)$ at $t_d$.\\
Comparing now $\eta$ with $\lambda$ (both have dimension of inverse time) we find over a wide range of potential strengths ($0.04e\leq V_{\rm A}\leq0.2e$) that the exponential separation on the mean-field level and the depletion on the many-body level correlate well with each other: an increasing Lyapunov exponent $\lambda$ with increasing $V_{\textrm{A}}$ goes hand in hand with an increasing $\eta$ (Fig.~\ref{fig:lambda_eta}). Up to a constant numerical factor ($\approx10$) $\eta$ follows $\lambda$ as a function of $V_{\textrm{A}}$. We note that both $\eta$ and $t_d$ are sensitive to the specifics of the fit function Eq.~\ref{eq:fit} which results in an uncertainty of the fit. The qualitative behavior remains, however, unchanged. Within the precision of the fit we find that $\eta$ is dependent on $N$ (which can be qualitatively seen in Fig.~\ref{fig:sn_d2_Npart}).\\
The association of $t_d$ with $t_e$ faces the difficulty that $t_d$, similar to $\eta$, is dependent on $N$. For example, for $N=10^4$ the onset of depletion $t_d\approx2t_0$ differs from $t_e\approx6t_0$ within the GPE. However, we observe that $t_d$ increases with increasing $N$ [see the variation as a function of the particle number $N$ in Fig.~\ref{fig:sn_d2_Npart} (b)]. We conjecture that the onset of depletion $t_d$ approaches the onset of wave chaos within the GPE, $t_e$, in the limit $N\rightarrow\infty$. To prove this conjecture it would be necessary to investigate $t_d$ over a wide range of $N$ which is, however, prevented by conceptual and numerical limitations: for small $N<1000$ the initial state shows deviations from the Thomas-Fermi limit (Fig.~\ref{fig:gs_density}) while large $N>10^5$ are numerically too demanding. The $N\rightarrow\infty$ limit remains therefore an open problem. However, Fig.~\ref{fig:sn_d2_Npart} demonstrates that $t_e$ is the upper limit for the depletion time for experimentally realized particle numbers of $N\lesssim10^5$.\\
For relatively small $N$ ($N=10^3$) the MCTDHB simulations are also feasible for $M=3$. Comparing $M=2$ and $M=3$, the threshold for depletion is only weakly dependent on the number of orbitals included: we obtain almost the same $t_d$ for $M=2$ and $M=3$.\\
Another important example is propagation in a disorder potential. We use the Gaussian correlated disorder potential for which we have observed a transition from algebraic to exponential localization as a function of the correlation length $\sigma$.\cite{BreColLudSchBur11} This transition has been first observed for the speckle potential \cite{SanCleLugBouShlAsp07,BilJosZuoCleSanBouyAsp08} and associated with its high-momentum cut-off in the Fourier spectrum.\cite{SanCleLugBouShlAsp07,LugAspSanDelGreMulMin09} We observe the same transition for Gaussian correlated disorder\cite{BreColLudSchBur11} where a high-momentum cutoff in the Fourier spectrum is absent.
We show the results for propagation in a disorder potential with parameters for which previously Anderson localization has been observed.\cite{BilJosZuoCleSanBouyAsp08} Averaging over several realizations of the disorder potential with $V_{\rm A}=0.2e$ and correlation length $\sigma=0.7\xi$ we obtain within the GPE an exponential increase which sets in several units of $t_0$ before the exponential increase for the periodic potential with $V_{\rm A}=0.2e$ and $l=0.54811l_0$ [see Fig.~\ref{fig:sn_d2_Npart} (b)]. In qualitative accord we observe that also $S_N(t)$ bends up earlier for the disorder potential than for the periodic potential [see Fig.~\ref{fig:sn_d2_Npart} (a)]. Our results suggest a destruction of the BEC as indicated by the occupation of excited modes during expansion in disorder potentials.\\
\begin{figure}[t]
	\centering
		\includegraphics[width=8.5cm]{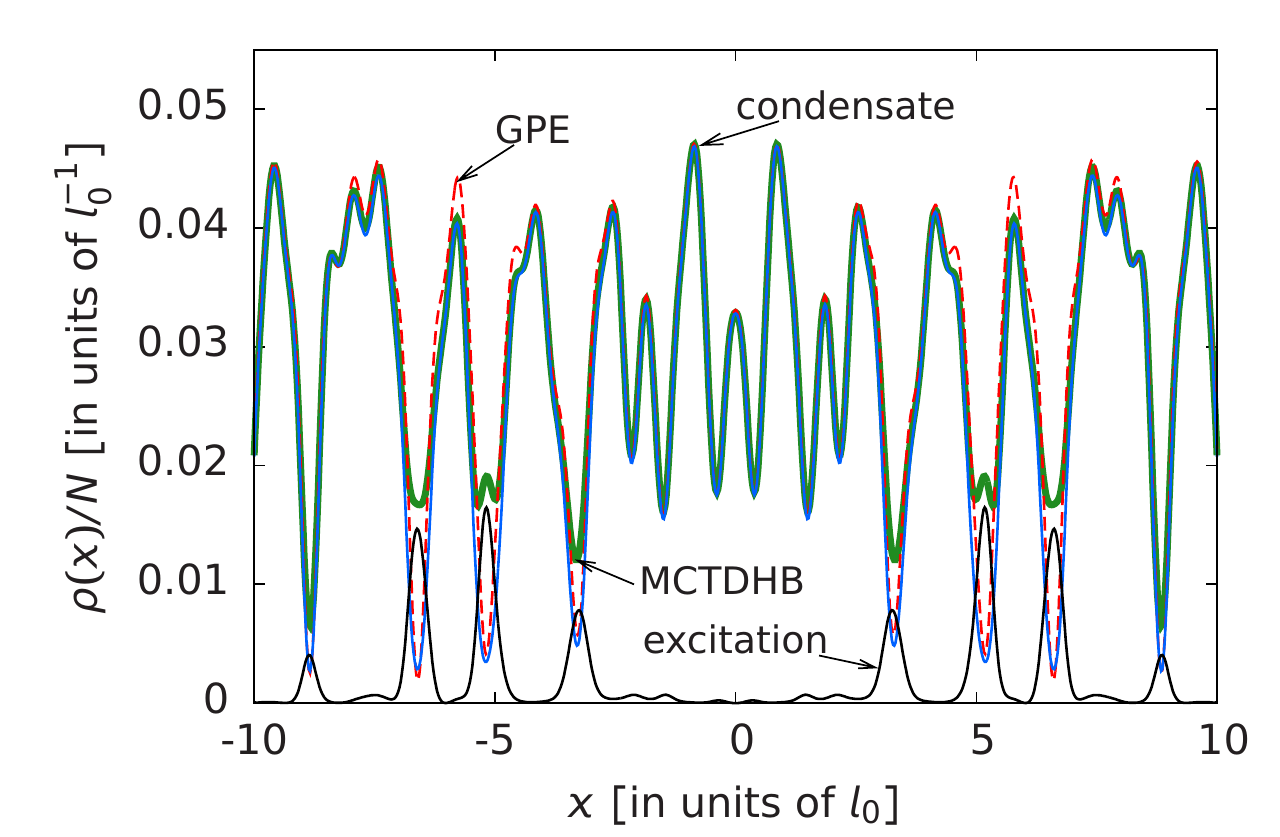}
	\caption{(Color online) Particle density at $t=4t_0$ within MCTDHB for $N=10^4$ (green thick line) and GPE (red dashed line). The condensate density is given by $n_1^{\rm NO}(t)|\Phi_{1}^{\rm NO}(x,t)|^2$ [blue (gray) line], the density of excited atoms is determined by $n_2^{\rm NO}(t)|\Phi_{2}^{\rm NO}(x,t)|^2$ (black line).}
	\label{fig:local_depletion}
\end{figure}
The onset of depletion of the condensate is mirrored in the fine scale oscillations of the density (Fig.~\ref{fig:local_depletion}). Substantial deviations within the GPE from the density obtained within the MCTDHB method emerge at different instants of time for different particle numbers. For $N=10^4$ deviations in the local fluctuations of the density emerge at $t\approx4t_0$ (see Fig.~\ref{fig:local_depletion}) monitored by $S_N(t)>0$. The occupation numbers are $n_1^{\rm NO}\approx0.96$ and $n_2^{\rm NO}\approx0.04$. While the condensed part [given by $n_1^{\rm NO}(t)|\Phi_{1}^{\rm NO}(x,t)|^2$] still closely follows the GPE prediction $|\psi(x,t)|^2$, the total density $\rho(x)$ shows smoothing of the local fluctuations near the center. This smoothing is due to excited atoms whose density partially fills in the local minima. For the system with $N=10^5$ the picture is very similar except that the occupation of the excited state is lower at $t=4t_0$, $n_2^{\rm NO}\approx0.003$ instead of $n_2^{\rm NO}\approx0.04$. The initially spatially localized excitations spread over the entire system with increasing time. One can expect the fine scale structure of the density of the full many-body system to strongly differ from the prediction of the GPE.\\
The discrepancies in the particle density go hand in hand with the breakdown of coherence as measured by the normalized two-particle correlation
function (Fig.~\ref{fig:g2_N104}).
In the regions of high density of excited atoms (near the local maxima of the second
natural orbital) the two-particle coherence is lost; ~$g^{(2)}$
strongly differs from $1$. The deviation of $g^{(2)}$ from unity indicates that the
many-body state is no longer representable by a product of a single complex-valued function. Consequently, the GPE ceases to be a valid description. This is a fingerprint of the emerging fragmentation of the many-body system.\\
For longer time intervals our MCTDHB calculations indicate a destruction  or at least a strong fragmentation of the condensate. 
For $t\gtrsim10t_0$, e.g, the occupation of both orbitals is approximately $50\%$ indicating that many more orbitals would be required for convergence. Nevertheless, current experiments indicate remarkable agreement with the prediction of the GPE for coarse-grained observables such as the width of the atom cloud or the average position (see e.g.~Ref.~\onlinecite{SchDreKruErtArlSacZakLew05,LyeFalModWieForIng05,CleVarHugAsp05,SanCleLugBouShlAsp07,BilJosZuoCleSanBouyAsp08,DriPolHitHul10}).
The width
\begin{equation}
\Delta x=\sqrt{\langle x^2\rangle-\langle x\rangle^2},
\end{equation}
where $\langle x^n \rangle=\int dx \rho(x,t)x^n$ is independent of wave chaos:\cite{BreColLudSchBur11} Even though two close wave functions $\psi_1(x,t)$ and $\psi_2(x,t)$ develop random local fluctuations, the width for both $\psi_1(x,t)$ and $\psi_2(x,t)$ agrees. If we now compare the prediction for the width within the GPE and within the MCTDHB method, we observe the same trend. While the fine scale structures of the wave function within MCTDHB have not fully converged for the small number of orbitals ($M\leq3$) included in the simulation, the coarse-grained distribution remains essentially unchanged compared to the GPE [Fig.~\ref{fig:xn_kn_average} (a)]. We thus expect that the time dependence of the width of the full many-body system is well accounted for by the GPE [Fig.~\ref{fig:xn_kn_average} (a)].
Despite its failure to account for the state entropy (Fig.~\ref{fig:sn_d2_threshold} and Fig.~\ref{fig:sn_d2_Npart}) and the coherence properties (Fig.~\ref{fig:g2_N104}), the GPE thus remains predictive in describing the expansion of a BEC in external potentials on longer time scales for coarse-grained observables, long after the random fluctuations prevent the prediction of fine scale structures in $\rho(x,t)$. Up to now, local small-scale fluctuations have not been investigated experimentally because of the difficulty of (sub) $\mu$m resolution. 
The same excellent agreement we observe for the average over momenta $k^2$ as accessible in time-of-flight experiments [see Fig.~\ref{fig:xn_kn_average} (b)]. The average over $k^2$ is determined via 
\begin{equation}
\langle k^2 \rangle=\int\; dk\; k^2 \tilde{\rho}(k,t)
\end{equation}
and is proportional to the kinetic energy per particle. The GPE thus reproduces the mean kinetic energy of a highly excited system despite its failure to account for breakdown of coherence, fragmentation, and small-scale fluctuations. The latter observation indicates that thermalization may be within the realm of the GPE despite its failure to account for two-body scattering which is key to any thermalization process.
\section{Conclusions}\label{sec:conc}
By comparing simulations within the Gross-Pitaevskii equation (GPE) and the multiconfigurational time dependent Hartree for bosons (MCTDHB) method we have uncovered that wave chaos in the GPE indicates depletion of the occupation of a BEC during expansion in the presence of weak external 1D potentials. We have checked that this connection holds for a large class of external potentials including a harmonic potential with short-ranged perturbation (not shown), an aperiodic potential with incommensurate frequencies, and disordered and periodic potentials explicitly discussed in this paper. This connection has far-reaching consequences: while the depletion and fragmentation process is an intrinsic many-body effect outside the realm of the GPE, the mean-field theory allows one to monitor its onset through the development of random local fluctuations. The measure for the random local fluctuations, $d^{(2)}(t)$, can be used to delimit the applicability of the GPE to approximate the many-body dynamics. On the many-body level the depletion process manifests itself through the loss of coherence as measured by deviations of $g^{(2)}$ from unity. We point out that the connection between wave chaos and depletion is unidirectional:  The presence of depletion on the many-body level does not necessarily imply the presence of wave chaos on the mean-field level. Similarly, the absence of wave chaos does not imply absence of depletion. Rather, for every system where we have found wave chaos within the GPE the occupation of the BEC abruptly decreases. Coarse-grained (``macroscopic") quantities become independent of random (``microscopic") fluctuations. Thus, wave chaos identifies a depletion process which eventually may lead to relaxation and thermalization (see e.g.~Ref.~\onlinecite{Sre94,RigDunOls08,MazSchSch08,CasClaRig11}).
The depletion process, the onset of which we have investigated, can be experimentally studied provided a sufficient spatial resolution is achieved. Observables include higher-order coherence, i.e. deviations of $g^{(2)}$ from unity as measured e.g.~in Ref.~\onlinecite{ManBucBet10}. It would be of considerable interest to verify experimentally our predictions by exploring the fine-scale fluctuations and coherence properties of expanding BECs in external potentials and thus gain deeper insight into the involved many-body effects.
\section*{Acknowledgments}
We thank Moritz Hiller, Fabian Lackner, Hans-Dieter Meyer, Kaspar Sakmann, and Peter Schlagheck for helpful discussions. This work was supported by the FWF doctoral program ``CoQuS". Calculations have been performed on the Vienna Scientific Cluster and the bwGrid. Financial support by the DFG is acknowledged.

\end{document}